\documentclass[11pt]{article}
\usepackage[a4paper,margin=2.6cm]{geometry}
\usepackage{amsmath,amssymb,amsthm}
\usepackage{booktabs}
\usepackage{hyperref}
\hypersetup{pdftitle={Unit: a type checker for dependent type theory}}

\newcommand{\sys}{\textsc{Unit}}
\newcommand{\kw}[1]{\texttt{#1}}
\newcommand{\U}{\mathsf{U}}

\title{\sys: a type checker for dependent type theory}
\author{Thierry Coquand}
\date{}

\newcount\stamphour \newcount\stampmin
\stamphour=\time \divide\stamphour by 60
\stampmin=\time \multiply\stamphour by 60 \advance\stampmin by -\stamphour \divide\stamphour by 60
\edef\stamptime{\the\year-\ifnum\month<10 0\fi\the\month-\ifnum\day<10 0\fi\the\day\space
 \ifnum\stamphour<10 0\fi\the\stamphour:\ifnum\stampmin<10 0\fi\the\stampmin}
\AddToHook{shipout/foreground}{\put(0.5\paperwidth,-\dimexpr\paperheight-0.45in\relax)%
 {\makebox[0pt]{\scriptsize\ttfamily\jobname.pdf\quad\stamptime}}}
\begin{document}
\maketitle

\begin{abstract}
We present \sys, a type checker for dependent type theory with a module
system in the tradition of \textsc{Automath} and a uniform treatment of
definitions. There is no unification, no instance search and no
tactics, and there is one uniform unfolding policy, without priorities
for particular definitions or choices adapted to the problem, so the
work of checking can be observed and compared with the informal proof. With \sys\ we have
checked a model of univalent type theory, synthetic algebraic geometry
(schemes, the \'etale topology, Severi--Brauer varieties), synthetic Stone duality with its
cohomology, quasi-compact quasi-separated schemes and their functors of
points, and results of commutative algebra, among them
finite free resolutions, Wiebe's theorem, Zariski's main theorem and the
local structure of \'etale algebras. The system is neutral as to logic:
most developments are constructive, but classical principles can be
assumed where they are wanted, as excluded middle is in the simplicial
set model. These developments show that a small, uniform system
supports substantial mathematics, and give concrete reasons to
reconsider the complexity routinely accepted in proof assistants.
\end{abstract}

% ======================================================================
\section{Introduction}

Large proofs are now checked in dependent type theory. Its basic ideas
go back to \textsc{Automath}: propositions as types,
proofs as terms checked in a context, and definitions that are not
expanded in the proof but expressed by convertibility. That definitions
belong to the calculus, and are unfolded by a reduction of their own, is
the point of de Bruijn's $\Lambda\Delta$~\cite{LD}, where a definition is
a declaration of the same kind as an abstraction and can occur anywhere
in a context. Here we investigate a particular combination: shared
representations of definitions, a uniform conversion strategy, and
reports of the definitions unfolded, compared with an accompanying
mathematical proof.

So far, proof systems have been designed for interaction with a human
user, and pure dependent type theory has been extended for this purpose:
for instance with unification and implicit arguments, type classes
or instance search, coercions, tactics, impredicative propositions and
definitional proof irrelevance. Each serves a purpose in large-scale
formalisation, but they make the work of checking harder to relate to
the proof being checked, which is what we want to study here, and they
are not necessary for representing mathematical arguments: the
developments of Section~\ref{sec:results} were written without
unification-based argument inference, instance search, coercions,
tactics or a proof-irrelevant sort; omitted arguments are only filled by
one-way matching (Section~\ref{sec:conv}). We
therefore work with a small dependent type theory with definitions,
modules and records, in which nothing is guessed and every unfolding of
a definition can be observed. In particular there are no tactics: a proof is the term that is
checked; and there are no placeholders to be refined interactively: a
file is checked as it is written.

This paper presents such a system, \sys\ (Section~\ref{sec:checker});
the checker and all the developments are available at
\url{https://github.com/coquand/unit}.
Its type theory has $\Pi$, $\Sigma$, records, data types with primitive
recursion, an identity type, and cumulative universes indexed by level
expressions. Its module system follows the segments of \textsc{Automath}:
a theorem is stated in a context, and to use it one gives an instance of
the context, the objects and the proofs of the hypotheses, in order; an
instance only binds the parameters, and the body is neither copied nor
checked again. Its treatment of definitions is uniform: a definition is
a declaration like the others, it keeps one identity wherever it is
used, and conversion tries to recognise two values by their names before
it opens anything, and then opens one definition at a time.

Nothing in the checker commits a development to constructive logic.
Excluded middle, choice or any other principle is a hypothesis like the
others, a parameter of the segment whose files use it
(Section~\ref{sec:checker}). Most of the developments below are
constructive because their mathematics is, but classical reasoning can
be used where it is wanted: the simplicial set instance of the model of
univalent type theory is checked with excluded middle for propositions,
and its parameter shows exactly which results depend on it.

This uniformity is not only a matter of simplicity. In \cite{Rem} the
author recalls that definitions are used in type theory differently from
the way Frege used them in the \emph{Begriffsschrift}~\cite{Frege}, and
that, on the first lemmas of the \emph{Begriffsschrift} translated into
\textsc{Automath} notation, type checking felt similar to what one does
when trying to understand a proof. We ask whether this holds beyond this
example. We consider detailed, precise informal
proofs that identify the named constructions and lemmas used and explain
the necessary openings of definitions. Ordinary texts often leave such
openings silent; making them explicit is part of the comparison, not an
assumption about all mathematical reading. We make two hypotheses about
these selected aspects of checking.
\begin{enumerate}
\item[(H1)] \emph{Checking can follow reading.} On a proof accompanied
  by such a text, a checker can recognise repeated uses of named
  constructions without unfolding them, and its necessary openings of
  definitions can be accounted for by the text or by a stated
  representation convention. This concerns recognition and unfolding;
  substitution, universe constraints, scope checking and argument
  recovery remain additional work of the checker.
\item[(H2)] \emph{Cost is a diagnostic.} When a step that is routine on
  paper is expensive to check, the report can help locate repeated
  decoding of a representation, a missing interface or intermediate
  statement, or unnecessary work in the checker. This is a heuristic,
  not an exhaustive diagnosis or a complexity bound: some proofs do
  compute.
\end{enumerate}
For (H2) to be usable the checker must not hide the cost: a checker with
clever unfolding heuristics can make a badly structured proof cheap, and
so lose the information. A uniform checker makes its work observable.
This should be relevant to the problem of proofs \emph{lost in
translation}~\cite{BCH}: a formal proof can be accepted without being the
proof of the text it claims to translate. Unfolding reports supply
evidence for investigating this correspondence. An unexplained opening
may be representational work within the same mathematical argument;
agreement of reports cannot exclude a different argument or an
incorrectly translated statement.

\paragraph{The method} (Section~\ref{sec:method}) uses this. It has
three parts.
(1)~The checker reports, for each declaration, how its comparisons were
settled and which definitions it opened. (2)~The informal proof is
written, and each step is classified by what a reader does: match
names, open a definition the text cites, or nothing. The definitions the
text cites can be given to the checker, which then reports the others.
(3)~For a declaration that costs more than its text suggests, the formal
proof is split into the steps of the informal proof, each step is
measured, and the expensive comparisons are traced to the part of the
checker that asks for them. The method can be applied to an existing
development, or while writing a new one.

\paragraph{The aim} (Section~\ref{sec:align}) is that the structure of a
formal proof be that of a detailed, precise informal proof: one
definition for each notion, one declaration for each step, the
intermediate claims written, ``by definition'' where a definition is
opened, and each fact that a text states together with a definition
proved once, as a lemma that holds by definition. Acceptance does not establish that a formal proof follows
a given text \cite{BCH}; the method supplies local evidence for
examining this question.

\paragraph{The claim.} \sys\ demonstrates that a small, uniform system
supports substantial mathematics. Its developments, more than thirty
thousand declarations, go from a model of univalent type theory to
schemes, the \'etale topology and Severi--Brauer varieties, to the comparison of sheaf and \v{C}ech
cohomology of compact Hausdorff spaces, and to finite free resolutions
in commutative algebra, with no unification, no instance
search, no tactics, and one uniform unfolding policy without
definition-specific priorities or adaptive choices, and each of them is
checked in under a minute (Table~\ref{tab:valid}). Each notion these
systems add solves a real problem; the developments show that the
problem can also be solved by writing, with named instances, explicit
structures and stated lemmas, and in the cases we examined
(Section~\ref{sec:results}) this brought the formal proof closer to its
text. This is a concrete reason to reconsider the complexity routinely
accepted in proof assistants for mathematics.

AI-assisted formalisation makes the question pressing. Much of that
complexity exists to spare a human author from writing what can be
inferred. When the formal text is written by a machine, explicit
arguments still cost something to generate, maintain and review, but
sparing a human the typing is no longer a decisive reason for
complexity in the proof language or in the checker; what matters is
that the result can be checked simply and read against the
mathematics. All the developments
reported here were written in this way. Existing proof assistants remain
valuable systems and libraries, but their architecture need not be the
inevitable model for what comes next: it is a reason to reconsider which
complexities are necessary in the proof language and in the checker.

\paragraph{Results} (Section~\ref{sec:results}).
Table~\ref{tab:totals} gives what checking opens in each development.

% ======================================================================
\section{The checker}\label{sec:checker}

\subsection{Units and segments}

A \emph{unit} is a declaration:
\[
\begin{array}{ll}
x\,(\Delta) : A & \text{a parameter}\\
x\,(\Delta) : A = t & \text{a definition}\\
\kw{data}\ D\,(\Delta) = c_1(\Delta_1)\mid\dots & \text{a data type}\\
\kw{level}\ a,\quad \kw{constraint}\ \psi & \text{levels and constraints}\\
\kw{module}\ M\,(\Delta)\,\{\,\vec u\,\},\quad \kw{record}\ R\,(\Delta)\,\{\,\vec u\,\}
  & \text{a segment}\\
\kw{module}\ N = M[\vec a] & \text{an instance}\\
\kw{opaque}\ f,\quad \kw{transparent}\ f & \text{the unfolding strategy}
\end{array}
\]
Terms are those of a dependent type theory with $\Pi$, $\Sigma$, an
identity type with its eliminator $J$, labelled sums with their
eliminators (\kw{split}), records, cumulative universes \kw{U}\,$l$
indexed by level expressions, and products over levels and over
constraints between levels~\cite{BCDE}.

A file, a module body, a record body and a local block are all lists of
units, checked in the same way: each unit in the context of the units
before it, and then added to it; a definition stays a definition, with
its body available for computation. The scope is sequential: a unit sees
only the units before it, never itself or a later one. This is the block
structure of \textsc{Automath}, and, as in $\Lambda\Delta$~\cite{LD}, a
context contains definitions as well as declarations, in any order. The
context after a unit does not depend on whether the unit checked: a
definition enters the context with its type and its body, not with its
proof of correctness. So each declaration can be checked separately,
with its own time budget, and one file can be checked while the others
are trusted, as a reader accepts a lemma and reads on. This serves
diagnosis only: a development is accepted when all its declarations
check, and the success of a declaration is relative to the context
before it, so that after a failure it is conditional on the failed
declaration.

A \emph{segment} is a named list of units; its parameters are its units
without a body. It is checked once, with its parameters unknown. An
instance $M[\vec a]$ checks each $a_i$ against the type of the $i$th
parameter with the earlier ones substituted; nothing of the body is
checked again. A module and a record are the same object and differ only
in what is given: an instance gives every parameter; a record type
$R[\vec p]$ gives those of the header, and \kw{new}\,$R[\vec p,\vec v]$
gives the others, its fields. Fixing a field, $R[\vec p][x:=e]$, gives
the type of the elements whose field $x$ is $e$. A hypothesis can be a parameter of a whole
development and be discharged by one instance; nothing is postulated.
The model of univalent type theory below assumes function
extensionality and univalence as two parameters, and one instance,
which checks two arguments, gives it under univalence alone.

\paragraph{Classical logic.} In the same way, \sys\ can be used for
classical mathematics. Excluded middle for propositions,
\[
\kw{lem} : (P : \U\ b) \to \kw{isProp}\ P \to P + (P\to\bot)\quad
\text{for every level } b,
\]
is the parameter of the segment \kw{Cl} of \texttt{core/CClassical.seg},
and the four files of the simplicial set instance that need it are
imported inside it.
It is used at one point. The fillers built constructively are uniform
along the face maps only. To make them uniform along the degeneracies
too, the Eilenberg--Zilber normalisation asks whether a filling problem
is degenerate. That is an equation between partial elements, a
proposition, and \kw{lem} decides it. The rest of the model, and the
instance on finite posets, do not use it. Excluded middle is assumed for
propositions only, so it is compatible with univalence; for all types
it would contradict it. The axiom of choice can be assumed too, as a parameter stating that for a set $X$ and a family of sets
$Y$, $\prod_{x:X}\lVert Y\,x\rVert\to\lVert\prod_{x:X}Y\,x\rVert$; in this
form it is also compatible with univalence. A classical development is written in the same
way, with \kw{lem} or a choice principle as a parameter of the whole
development; nothing in the checker changes. As in Escard\'o's
\textsc{TypeTopology}~\cite{TT}, where such assumptions are hypotheses
of the theorems that use them, the assumptions of a development are
explicit, and a file can only be used where they are given
(Section~\ref{sec:results}).

Data types are strictly positive. There is no recursion by name: a
function out of a data type is a \kw{split}, and a branch names, after
its pattern variables, the recursive results, which are hypotheses like
any other:
\begin{verbatim}
  data N = zero | suc (n : N) ;
  plus (n : N) : N -> N = split { zero -> n ; suc m | r -> suc r } ;
\end{verbatim}
This is primitive recursion at higher types, as in G\"odel's system T,
written as the inductive case expressions of Burstall~\cite{Burstall},
where termination follows from the scope rules.

\subsection{Structures without instance search}\label{sec:structures}

Proof assistants commonly find structure arguments by instance
synthesis, or solve unification problems on record projections by
canonical structures; both choose automatically an argument that could
be given explicitly.
\sys\ has neither. A mathematical structure is a record, generic
mathematics about it is a module whose parameter is an element of the
record, and \kw{open} gives the operations and laws short names:
\begin{verbatim}
  record CRing {
    R : U a ;  z : R ;  o : R ;
    add : R -> R -> R ;  mul : R -> R -> R ;  neg : R -> R ;
    addC (x y : R) : Id R (add x y) (add y x) ;
    addZ (x : R) : Id R (add x z) x ;  ...
  } ;
  module RL (Rg : CRing) {
    open Rg ;
    unit (x : R) : U a = (y : R) * Id R (mul x y) o ;
    zl (x : R) : Id R (add z x) x = trans (addC z x) (addZ x) ;
    ...
  } ;
\end{verbatim}
The ring is given once, for a whole development, as a text says ``let
$R$ be a commutative ring'' and then uses its operations; an instance
\kw{RL[Rg]} makes all the results available for a particular ring. The
choice is explicit, but it is made once, not at each step.

The difficulty that instance search addresses arises when a proof moves
among many related structures: a ring, its additive group, a module over
it, a ring of polynomials. With explicit structures each passage is a
named construction, and the same structure reached by two constructions
must be identified explicitly. In the cohomology of synthetic Stone
duality, the groups of cocycles are both kernels of homomorphisms and
groups of the \v{C}ech complex; statements written with one name and
used with the other showed in the report as repeated openings (one
declaration caused 3708 unfoldings), and stating the types in the
terms of one construction removed them (447). In the developments of
Section~\ref{sec:results}, records, parameterised modules and \kw{open}
have sufficed to organise the structures without instance search, and
supplying structure arguments at the level of a development has
introduced little repetition.

\subsection{Definitions as closures}

Evaluation maps terms to values in an environment. Following
$\Lambda\Delta$, a definition is not replaced by its body when it is
declared; it is unfolded only where its body is needed. The use of
closures comes from~\cite{C96,Wang}, and was used in the implementation
of \texttt{cubicaltt}~\cite{cubicaltt}, which also introduced
\kw{opaque} and \kw{transparent}. A use of a definition $f$ evaluates
to the closure $\langle f;\rho\rangle$ of the unit and the environment
$\rho$ where it is declared. The \emph{arguments} of a use are the values
in $\rho$ of all the variables and levels in scope at the declaration. A
member $M[\vec a].f$ of an instance is $\langle f;\rho,\vec a\rangle$.
Nothing is copied: $f$ has one identity wherever it is used.

\kw{opaque f} declares that from this point on $f$ is not unfolded, and
\kw{transparent f} undoes it.
The strategy is a parameter of the run:
evaluation builds the same values whatever the strategy. Making a
definition opaque can only make fewer terms convertible, so an
\kw{opaque} declaration never makes an incorrect proof correct. Its use
is the mathematical one: once a construction is characterised by lemmas,
the rest of the development uses the lemmas.

\subsection{Conversion}\label{sec:conv}

Definitional equality is generated by $\beta$ (for functions, pairs,
levels, projections, \kw{split} on a constructor and $J$ on reflexivity),
the unfolding of definitions, and $\eta$ for functions and pairs. A data
type and a \kw{split} are constants, as a definition is in
$\Lambda\Delta$: a data type is applied to its parameters, and a
\kw{split} is lifted over the whole context in which it is written, as
local pattern-matching functions usually are. A \kw{split} computes
only on a constructor, and two occurrences of the same constant are
equal when their arguments are, an argument that a branch does not use
being still an argument. This is part of the declarative equality, not a
restriction of the algorithm: the algorithm is meant to decide exactly
this equality, although we give no proof that it does
(Section~\ref{sec:open}). Two values are compared as follows.
\begin{enumerate}
\item \emph{By names.} They are equal if they are strictly equal, that
  is, equal as represented, two closures $\langle f;\rho\rangle$,
  $\langle f;\rho'\rangle$ being strictly equal when $\rho$ and $\rho'$
  are; or if they are eliminations of the same definition $f$ with
  convertible arguments and spines. The common head is not unfolded;
  the arguments are compared by conversion, from stage 1, and this
  may unfold definitions inside them: comparing $f(g(a))$ with $f(b)$
  may open $g$, never $f$.
\item \emph{By head steps.} Otherwise the $\beta$-redexes at the head are
  reduced, without unfolding. If a definition blocks the head of one side
  only, that side is brought to weak head normal form. If both heads are
  blocked by definitions, the one declared later is opened by one
  $\delta$-step. After each step, strict equality is tried.
\item \emph{By components.} When no head is blocked, the outer forms are
  compared and their components are compared recursively, each from
  stage 1.
\end{enumerate}
Stage 1 is a reader who recognises the same lemma at the same arguments;
stage 2 is opening a definition, the $\delta$-step of $\Lambda\Delta$.
Stage 1 is essential: comparing arguments only strictly, and otherwise
unfolding, takes more than 30~s on each file of the model tried, against
1.5--4~s. The strategy is uniform: there is no other choice of what to open, and
no heuristic depending on the definitions. Sequential scope ensures that
an earlier declaration's body cannot refer to a later declaration. It
does not impose this order on instantiated values: if $f(x):=x$ is
declared before $g:=a$, then $f(g)$ unfolds to $g$. The later-first rule
is a fixed policy, not a completeness argument. Two heads can also
reduce to a common expression. If $h:=a$ and then $k:=a$, comparison
proceeds as
\[
  h\mathrel{\sim}k\quad\longmapsto\quad
  h\mathrel{\sim}a\quad\longmapsto\quad a\mathrel{\sim}a.
\]
Accordingly, two \kw{split} values, like two data types, are compared by
name: by the identity of their code and their convertible environment
arguments; their branches and motives are never compared;
two record types are equal when they come from the same declaration with
equal parameters, and their fields are never compared.

\paragraph{The expected type is opened only for introductions.} An elimination (an application, a
projection) is checked by inferring its type and comparing it with the
expected type as it is; the expected type is unfolded only when the term
is an introduction (a $\lambda$, a pair) and its outer form is needed.
A reader does not open the statement of a lemma to check that the lemma
has been applied.

\paragraph{Notation.} Omitted arguments are filled by one-way matching
of the inferred types of the explicit arguments against the domains,
never by unification; two different solutions for a hole are an error.
A solution is written back as a term in the current scope. The names in
scope are recognised by a restriction of stage~1 that opens nothing at
any depth: strict equality, or the same definition at arguments compared
in the same way, with no head step. A reader who writes an omitted
argument recognises names and does not compute. Only when no name in
scope denotes a definition occurring in the solution is that definition
replaced by its body, which is then written. Equational chains
\kw{calc}, \kw{open} and \kw{import} are translated before checking.

% ======================================================================
\section{The method}\label{sec:method}

\paragraph{The measurement.} With the option \kw{--opened}, \sys\ reports
for each declaration the number of comparisons (including subproblems)
settled at stage 1, by arguments (stage 1 with a recursive comparison of
arguments), and by components (stage 3), and the definitions unfolded
during its check, with multiplicities. We distinguish the
\emph{unfoldings}, each $\delta$-step being one, from the distinct
definitions opened: \kw{opened 2: Frege.her*3 Frege.Rplus*2} is five unfoldings of
two definitions. The unfolding of a data type to
its constructors is not counted: it is reading a declaration, not
opening a construction. The counters are active while a declaration's
checking result is evaluated, including argument matching and writing
back inferred arguments when these are forced in that interval. They
are not restricted to the final comparison of the inferred and expected
types. Parsing and name resolution are outside this interval; diagnostic
printing after the result is obtained is also outside it. With lazy
evaluation, shared work is attributed to the declaration that forces it.
The reported times are CPU times for these intervals. Unfolding counts
describe this instrumented execution, not all operations of the checker
and not a machine-independent cost of the proof.

\paragraph{What the text opens.} In HOL every use of a definition is an
explicit step, a rewrite with its defining theorem; in \textsc{Automath}
and in \sys\ the uses of a definition are silent. Objective type
theory~\cite{OTT} does the same as HOL for computation rules: $\beta$ and
the computation of the identity eliminator hold only propositionally,
with explicit witnesses, so each use is a transport written in the
term. An informal text is in
between: it says ``by definition of $X$'' where it uses a definition,
and says nothing at steps it regards as obvious. Frege had no
definitional equality for defining notions: in the
\emph{Begriffsschrift}~\cite{Frege} a definition is a judgement of
identity of content, and each use of it is a step citing it.
Each step of a text is classified by what a reader does:
\begin{center}
\begin{tabular}{lll}
\toprule
step of the text & the reader & the checker should open\\
\midrule
by Lemma $L$ & matches names & nothing\\
by definition of $X$ & opens $X$ & $X$\\
without comment & nothing & nothing\\
\bottomrule
\end{tabular}
\end{center}
With \kw{--opens F}, where each line of \kw{F} has the form
\kw{name: $d_1$ \dots\ $d_k$}, \sys\ prints for the declaration
\kw{name} the definitions it opened other than $d_1,\dots,d_k$. Each is
a use not listed in the supplied annotation. It may implement a
representational convention or an ordinary silent use of a definition;
its origin must be investigated before drawing a conclusion about the
mathematical argument.

\paragraph{Reading a declaration.} For a declaration that is slow, or
opens many more definitions than its text cites:
\begin{enumerate}
\item write its informal proof;
\item say how a reader checks each step: by a lemma, by a definition, or
  without comment;
\item split the formal proof into these steps, one lemma for each step
  with its intermediate points written, and measure each;
\item for the step that costs more than its text, find the comparisons
  that cost, and the part of the checker that asks for them (a traced
  version prints each comparison costing more than a threshold with its
  caller: subsumption, a \kw{refl}, the agreement of two solutions of a
  hole, the writing back of a solution, the arguments of the same
  definition, a \kw{split}, a locked definition).
\end{enumerate}
In these experiments the reports led to sharing one name for a construction, and to stating an interface or an
intermediate identification once at the generic level. These are
possible explanations and repairs, not an exhaustive classification of
unexplained unfoldings.

\paragraph{Writing.} The same method applies while writing, and gives
rules: write the informal proof first; one name for each notion (no
alias for something that has a name, pass a structure rather than its
components, name the components of a construction once where it is
defined); state the facts ``by definition'' once, by \kw{refl}, next to
the definition, and lock the construction after its characterising
lemmas; write the intermediate point of a step that the text proves by
two definitions; give accessors for the parts of elements rather than
chains of projections; after each declaration, compare what it opened
with what its text cites.

% ======================================================================
\section{First results}\label{sec:results}

\subsection{Small examples}

\paragraph{Sharing and the lock.} On
\begin{verbatim}
module Mon (A : U a) (e : A) (op : A -> A -> A) {
  twice (x : A) : A = op x x ; } ;
module NM = Mon[N, zero, plus] ;
\end{verbatim}
the report is:
\begin{center}
\begin{tabular}{lll}
\toprule
declaration & statement & opened\\
\midrule
\kw{twiceUnfold} & \kw{NM.twice x = plus x x} & \kw{Mon.twice}\\
\kw{byName} & \kw{Mon[N,zero,plus].twice x = NM.twice x} & nothing\\
\multicolumn{3}{l}{\kw{opaque Mon.twice}}\\
\kw{viaLemma} & \kw{NM.twice one = plus one one}, by \kw{twiceUnfold} & nothing\\
\multicolumn{3}{l}{\kw{transparent Mon.twice}}\\
\kw{compAgain} & \kw{NM.twice one = suc one}, by \kw{refl} & \kw{Mon.twice}, \kw{plus}, \kw{one}\\
\bottomrule
\end{tabular}
\end{center}
The same instance reached by two paths is recognised without opening
\kw{twice}; after the lock, the lemma \kw{twiceUnfold} is used instead
of the definition.

\paragraph{Frege's transitive closure.} The files
\texttt{examples/frege.seg} and \texttt{examples/frege.opens} of the
repository follow the first lemmas of the
\emph{Begriffsschrift}~\cite{Frege} on the transitive closure of a
relation, as in~\cite[\S2.4.3]{Rem}. The context, a type $A$ and a
relation $R$, is a segment:
\begin{verbatim}
module Frege (A : U a) (R : A -> A -> U a) {
  her : [b] (A -> U b) -> U (a \/ b) =
    <b> \ P -> (x : A) -> P x -> (y : A) -> R x y -> P y ;
  Rplus (x y : A) : U (a+) =
    (P : A -> U a) -> her @a P -> ((z : A) -> R x z -> P z) -> P y ;
  lem1 (x : A) : her @(a+) (Rplus x) =
    \ y h1 z h2 P h3 h4 -> h3 y (h1 P h3 h4) z h2 ;
  her-step (P : A -> U (a+)) (hP : her @(a+) P) (x : A) (px : P x)
           (y : A) (r : R x y) : P y = hP x px y r ;
  lem2 (x y z : A) (h1 : Rplus x y) (h2 : Rplus y z) : Rplus x z =
    \ P h3 h4 -> h2 P h3 (\ u r ->
      her-step (Rplus x) (lem1 x) y h1 u r P h3 h4) ;
  lem3 (x y z w : A) (h1 : Rplus x y) (h2 : Rplus y z) (h3 : Rplus z w)
    : Rplus x w = lem2 x z w (lem2 x y z h1 h2) h3 ;
  base (x y : A) (r : R x y) : Rplus x y = \ P h3 h4 -> h4 y r ; } ;
\end{verbatim}
It is instantiated, \kw{module FS = Frege[N, Succ]}, at the successor
relation \kw{Succ x y = Id N (suc x) y}, and two declarations prove that
$3$ and $2$ are in the closure of $0$: \kw{zero-three} by
\kw{FS.lem3} and \kw{FS.base}, and \kw{zero-two} directly, as a
function of \kw{P}, \kw{h3} and \kw{h4} that applies \kw{h3} and
\kw{h4} to the two steps.
Each line of \texttt{frege.opens} lists what the text opens at a step,
and \kw{--opens} reports the rest:
\begin{center}
\begin{tabular}{lll}
\toprule
declaration & the text opens & also opened\\
\midrule
\kw{lem1} & \kw{her}, \kw{Rplus} & nothing\\
\kw{her-step} & \kw{her} & nothing\\
\kw{lem2} & \kw{Rplus} & nothing\\
\kw{lem3} & nothing & nothing\\
\kw{base} & \kw{Rplus} & nothing\\
\kw{zero-three} & \kw{Succ} & \kw{one}, \kw{two}, \kw{three}\\
\kw{zero-two} & \kw{Rplus}, \kw{her}, \kw{Succ} & \kw{one}, \kw{two}\\
\bottomrule
\end{tabular}
\end{center}
The definition of hereditary is stated once, as the rule
\kw{her-step}, so that transitivity does not cite it; \kw{lem3} composes
two instances of transitivity by names, as the text does; through the
lemmas, \kw{zero-three} opens nothing of \kw{Frege}, while the direct
proof \kw{zero-two} opens the closure and heredity, where a text would
say ``by definition''. The numerals opened are a representation
convention: a reader takes $1$ to be the successor of $0$. An earlier
version of the checker also reported \kw{her} in \kw{lem2}, asked for by
no comparison. Step~4 of the method traced it to the checker: a
variable bound by $\lambda$, here \kw{h3 : her @a P}, is tested for
having a record type, and the test put its type in weak head normal
form. This is the third case of~(H2), unnecessary work in the checker;
the test is now not counted as an opening, which also removed
167\,702 unfoldings from the model of Table~\ref{tab:totals}. Finally,
Frege's proof of transitivity takes for $P$ the closure
\kw{Rplus x} itself, which is in \kw{U (a+)} and not in \kw{U a}; the
formal proof keeps the given $P$ and uses \kw{lem1} at the instance
$(x,y,h_1)$, as the comment in the file says. This is a change of
argument that no unfolding report can detect, made inspectable by
the accompanying explanation.

\subsection{The developments}

The method was applied to the following developments.
\begin{itemize}
\item A model of univalent type theory over a base category with an
  interval, with finite posets and simplicial sets as instances, and the
  localisation at a modality of Sattler (18\,051 declarations).
\item Synthetic algebraic geometry \cite{SAG}: schemes, the \'etale
  topology and Severi--Brauer varieties, up to the theorem that a
  Severi--Brauer variety with a point is projective space, from the
  axioms and a left exact sheafification for the \'etale
  topology (15\,770 declarations).
\item Synthetic Stone duality~\cite{SSD}: Stone spaces and countably
  presented Boolean algebras up to its Theorem~4.18, and sequential
  colimits and first cohomology up to its Theorem~5.16 (3310 and 4466
  declarations in its two parts, which share some files).
\item Cohomology in synthetic Stone duality~\cite{CCA}, on top of the
  previous one: the cohomology of compact Hausdorff spaces, up to the
  comparison of $H^n$ with \v{C}ech cohomology as a natural isomorphism
  of groups (8564 declarations with the files it reuses).
\item Commutative algebra includes finite free resolutions~\cite{CQ},
  \S\S2--8: McCoy's theorems, the Buchsbaum--Eisenbud structure theorem
  and Hilbert--Burch, also without bases. Other developments cover
  Koszul and true grade and Wiebe's
  theorem~\cite{CT}; local-global principles in depth~2 and Sharpe's
  theorem~\cite{CL}; the Dedekind--Mertens lemma; gcd domains and
  polynomial rings over them, using direct content induction; and the example of the Zariski main
  theorem of~\cite{ACL}, done by \v{C}ech cohomology; Jacobian
  regularity and the flatness of $A[X,Y][1/J]/(F,G)$ over any
  commutative ring. These share a library of polynomials, determinants
  and exterior algebra (6752 to 14\,190 declarations in six parts).
\item Quasi-compact quasi-separated schemes as locally ringed
  lattices~\cite{Zeu}: $\operatorname{Spec}$ and its universal property,
  gluing, the line with doubled origin and the projective line (qcqs and
  not affine), and the affine criterion; the functor of
  points~\cite{ZH}, fully faithful, with its essential image, giving an
  equivalence between the univalent categories of qcqs schemes as
  locally ringed lattices and as functors, also relative to a base ring;
  and Serre's affineness criterion (16\,047 declarations).
\item Zariski's main theorem in the form of~\cite{ACL}, over any
  commutative ring and in every dimension, by lifting finite sections.
  Its application is the local structure of \'etale algebras: an \'etale
  algebra over a residually discrete local ring is standard \'etale near
  a prime above the maximal ideal. It is checked with ZMT proved, and
  for a flat unramified algebra by a flatness transfer theorem
  (21\,057 and 7076 declarations). The field case, the structure of
  finite \'etale algebras over a discrete field, is still a hypothesis.
\end{itemize}
The gcd-domain development now follows the direct content induction of
Mines--Richman--Ruitenburg (Chapter~IV, \S4, Lemma~4.3), removing the
constant term to match coefficient-list order. It proves
$d\mid\text{coefficients}(fg)\Rightarrow d\mid c(f)c(g)$, from which
Gauss's lemma and the domain property of $R[X]$ follow. Its entry point
no longer imports Dedekind--Mertens. The proof and its correspondence
with the checked declarations are described in
\texttt{doc/gcd-polynomial.tex} and \texttt{doc/zmt-gcd-cech.tex}.
In the first two, about half of the declarations open no definition at
all (10\,026 and 9072). Stone duality, its cohomology and the later
developments were written from the start by the
rules of Section~\ref{sec:method}, comparing after each file what its
declarations open with what their texts cite.

The finite free resolutions are checked under two sets of assumptions.
\texttt{sag/cffr} states the results for matrices and assumes function
extensionality, univalence for propositions (equivalently, propositional
extensionality; it is used only for quotients) and replacement, not
univalence. \texttt{sag/cffru} states them for finite free modules
without bases and uses univalence, through the structure identity
principle for modules: the structures of the Buchsbaum--Eisenbud
theorem form a contractible type, the regular gcds of two elements form
an $R^\times$-torsor, and the determinant of a resolution is a line,
with no sign to choose. The univalent lemmas of the shared library (the
structure identity principle for algebras, the uniqueness of the
exterior algebra, torsors and cocycles) are in files of their own, which
\texttt{sag/cffr} does not import.

Table~\ref{tab:valid} gives the check of the developments by the
checker described here.
\begin{table}[ht]
\centering
\small
\begin{tabular}{llrrr}
\toprule
development & entry point & declarations & failed & time\\
\midrule
model of univalent type theory & \texttt{core} & 18\,051 & 0 & 46.5 s\\
schemes, \'etale, Severi--Brauer & \texttt{sag/chatelet} & 15\,770 & 0 & 7.3 s\\
Stone duality & \texttt{sag/stone} & 3\,310 & 0 & 1.0 s\\
sequential colimits, $H^1$ & \texttt{sag/stoneseq} & 4\,466 & 0 & 1.4 s\\
cohomology, \v{C}ech comparison & \texttt{sag/ccacoh} & 8\,564 & 0 & 4.1 s\\
finite free resolutions & \texttt{sag/cffr} & 9\,622 & 0 & 5.6 s\\
\quad without bases & \texttt{sag/cffru} & 14\,190 & 0 & 8.5 s\\
grade, Wiebe's theorem & \texttt{sag/wiebe} & 8\,354 & 0 & 3.2 s\\
depth 2, Sharpe's theorem & \texttt{sag/depth2} & 8\,386 & 0 & 3.4 s\\
gcd domains, Zariski main theorem & \texttt{sag/acl-zmt} & 7\,199 & 0 & 6.2 s\\
Jacobian, flatness & \texttt{sag/etale-flat} & 6\,752 & 0 & 2.5 s\\
qcqs schemes, functor of points & \texttt{sag/qcqs} & 16\,047 & 0 & 20.9 s\\
Zariski main theorem, \'etale & \texttt{sag/etale-local-zmt} & 21\,057 & 0 & 14.4 s\\
flat and unramified & \texttt{sag/etale-flat-unramified} & 7\,076 & 0 & 2.6 s\\
\bottomrule
\end{tabular}
\caption{The developments checked by the current checker: declarations
  (with the files they use), failures, and CPU time of the whole run,
  loading included; single runs under macOS on arm64, GHC 9.6.7,
  \texttt{-O1}.}
\label{tab:valid}
\end{table}

\subsection{Totals}

Table~\ref{tab:totals} gives, for each development, what checking
opens: the declarations that open no definition, the unfoldings in
all, the declarations with more than 500 unfoldings, and the largest
number of unfoldings in one declaration.
\begin{table}[ht]
\centering
\small
\setlength{\tabcolsep}{4pt}
\begin{tabular}{lrrrrr}
\toprule
development & declarations & open nothing & unfoldings & above 500 & largest\\
\midrule
model of univalent type theory & 18\,051 & 10\,026 & 1\,858\,930 & 601 & 171\,693\\
schemes, \'etale, Severi--Brauer & 15\,770 & 9\,072 & 255\,647 & 61 & 3\,130\\
Stone duality & 3\,310 & 1\,785 & 44\,622 & 5 & 953\\
sequential colimits, $H^1$ & 4\,466 & 2\,443 & 56\,966 & 5 & 953\\
cohomology, \v{C}ech comparison & 8\,564 & 4\,917 & 102\,630 & 5 & 953\\
finite free resolutions & 9\,622 & 5\,492 & 134\,666 & 14 & 1\,700\\
\quad without bases & 14\,190 & 7\,866 & 210\,940 & 24 & 1\,987\\
grade, Wiebe's theorem & 8\,354 & 4\,705 & 107\,787 & 10 & 1\,360\\
depth 2, Sharpe's theorem & 8\,386 & 4\,824 & 107\,598 & 10 & 1\,360\\
gcd domains, Zariski main theorem & 7\,199 & 3\,882 & 1\,968\,137 & 49 & 407\,664\\
Jacobian, flatness & 6\,752 & 4\,817 & 73\,190 & 9 & 8\,910\\
qcqs schemes, functor of points & 16\,047 & 9\,835 & 282\,932 & 87 & 8\,280\\
Zariski main theorem, \'etale & 21\,057 & 12\,378 & 414\,805 & 72 & 25\,673\\
flat and unramified & 7\,076 & 3\,954 & 89\,940 & 8 & 1\,987\\
\bottomrule
\end{tabular}
\caption{Unfoldings in checking the developments of
  Table~\ref{tab:valid} (\kw{--opened}), with the files they use;
  each $\delta$-step counts once, so a definition opened $n$ times
  counts $n$ times.}
\label{tab:totals}
\end{table}
Stone duality and its cohomology, written by the rules of
Section~\ref{sec:method}, have five declarations with more than 500
unfoldings, all on Boolean algebras and shared by all three. The model was written
before these rules. In most of its files what is opened is the decoding
of element types: a text writes an element of an extension $\Gamma.A$
as a pair $(\gamma,a)$ without comment, while the checker opens the code
of the extension at each projection. Its largest declaration opens one definition
140\,464 times; such declarations have not yet been read against their
texts. The exterior algebra and determinants of the algebraic
geometry (2122 declarations with the files they use) cause 29\,801
unfoldings; the largest declaration causes 1360, almost all of them (1255)
of two aliases, \kw{Alg} and \kw{Ext}, of notions that have names.
In the commutative algebra, the largest numbers come from the example of
the Zariski main theorem: its polynomial identities in five variables
are checked by computing normal forms, and each opens the multiplication
and addition of the normal forms some $10^5$ times (407\,664 for the
largest). This is computation stated as such in the text, not a
definition opened silently. In the proof of Zariski's main theorem the
largest declaration, 25\,673 unfoldings, compares two presentations of
the same polynomial ring, and 24\,044 of its unfoldings are of the two
names of that ring. This is a diagnostic in the sense of (H2): the
remedy is one name for the ring, not a different checker.

% ======================================================================
\section{Alignment}\label{sec:align}

That a formal proof is accepted does not establish that it is the
proof of the text. This is the point of \cite{BCH}, who show that the
formalisation accompanying an announced proof of blow-up for the
Navier--Stokes equations does not follow the text (some lemmas are proved
in a different, weaker form), and argue that a faithful translation of a
mathematical text cannot in general be guaranteed by a computation.

The aim here is to make selected aspects of the correspondence
inspectable: the named constructions, applications of lemmas and
openings of definitions in a formal proof are compared with a detailed,
precise informal proof. The following table describes the intended
organisation, rather than a criterion deciding fidelity:
\begin{center}
\begin{tabular}{ll}
\toprule
informal proof & formal proof\\
\midrule
a notion, with its name & one definition, used under that name\\
a lemma, a step & a declaration\\
an intermediate claim & a point written in the chain\\
``by Lemma $L$'' & an application of $L$, checked by names\\
``by definition of $X$'' & an opening of $X$, declared\\
a fact stated with a definition & a lemma proved by \kw{refl}\\
\bottomrule
\end{tabular}
\end{center}
The checker exposes computational work whose mathematical justification
or representational origin must be explained. An opening absent from
the annotation can arise from a second name for a notion, decoding of a
representation, an unstated intermediate identification, or an ordinary
use of a definition left silent by the text. Each requires examination;
none by itself proves that the arguments differ. Conversely, agreement
of unfolding reports cannot establish that the statements or arguments
correspond. A human reader must still compare both. The report supplies
local evidence for that comparison, maintained while the proof is
written.

A formal proof is well written, in this sense, when its structure is
that of a detailed informal proof, and its openings can be explained by
that text and its representation conventions. The informal proof this
asks for is longer than
usual: it has the explicit connections between convertible notions,
each a lemma proved by \kw{refl}, checked once, and free to use.

The repository has such accompanying texts for parts of the larger
developments. \url{doc/chatelet.pdf} gives the theorem on Severi--Brauer
varieties as checked in \url{sag/chatelet}, with the changes made to the source
note and a correspondence with the formal declarations;
\url{doc/horrocks-divisor.pdf}, \url{doc/horrocks-determinant.pdf}
and \url{doc/horrocks-gluing.pdf} give three proofs of rank-one
Horrocks, the first with a correspondence to \url{sag/horrocks}, the
last organised as the formal proof suggested; and
\url{doc/finite-projective-flat.pdf} gives a correspondence for the
characterisation of finite projective modules as finitely presented
flat modules.

% ======================================================================
\section{Related work}\label{sec:related}

\paragraph{Definitions in the calculus.} De Bruijn's
$\Lambda\Delta$~\cite{LD} makes definitions part of the typed
$\lambda$-calculus, with a reduction of their own, and his
\emph{mathematical vernacular}~\cite{MV} is a language meant to be close
to the way mathematics is written. The block structure of
\textsc{Automath} is that of \sys. The theories of \textsc{Half}
\cite{Ced} are lists of declarations; Wang~\cite{Wang} implements a type
theory with definitions as closures, which \sys\ follows.

\paragraph{Controlling unfolding.} Implementations of type theory decide
what to unfold by heuristics: the kernel described in~\cite{Lean4} unfolds
first the definition of greater height, which is close to our rule of
opening the later definition, and other systems use priorities set by
the user. Opaque definitions, and the controlled unfolding of
\cite{CU}, make unfolding a matter of the theory rather than of the
implementation. Here the strategy is kept uniform, and the resulting
reports are used to compare unfolding with an accompanying text.

\paragraph{Computation as explicit terms.} In objective type
theory~\cite{OTT} the only judgemental equality is syntactic identity;
the computation rules of $\Pi$ and of the identity type are
propositional equalities with explicit witnesses, and type checking is
quadratic. Its rules are interpretable in \sys, each witness becoming
\kw{refl} (the unbased eliminator being defined from the based one),
but not conversely: \sys\ has more equalities, hence fewer models, and
does not inherit the quadratic bound. The two make opposite choices for
the same information. Objective type theory records each use of a
computation in the proof term; \sys\ leaves the term unchanged, lets it
serve at a definitionally equal type, and reports which definitions the
comparison needed.

\paragraph{Structured formal proofs.} Mizar, Isar~\cite{Isar} and
Lamport's structured proofs~\cite{Lamport} aim at formal texts whose
structure is that of an informal proof; this is the aim of
Section~\ref{sec:align}, which adds a measurement of what each step
opens. De Bruijn's
estimate of the ratio between the length of a formal text and that of
the informal one~\cite{Survey} compares sizes, not structures.

\paragraph{Interfaces and constructions.} In the formalisation of
schemes in Lean~\cite{SiL}, the algebraic gluing theorem was first
proved for particular constructions of the localisation of a ring. The
scheme construction then produced iterated localisations
$R[1/f][1/g]$, canonically isomorphic to $R[1/fg]$ but not equal to it,
and the canonical identifications that a text makes silently had to be
supplied as comparison maps and commuting diagrams. The repair was to
state the theorem for any algebra satisfying a localisation predicate,
which the authors took to be three elementary conditions due to
Strickland, equivalent to the universal property. This is the second
case of~(H2), a missing interface, and the repair is the usual
mathematical one: construct a localisation once, prove its universal
property, and state everything else for an arbitrary algebra with that
property. The localisations of \url{sag/affine} and \url{sag/closed} are
written in this way. A localisation is an algebra with the universal
property; one construction, adjoining $y$ with $fy=1$, shows that it
exists; Strickland's conditions are derived as lemmas (that every
element is a fraction, and that the kernel of $A\to A[1/f]$ consists of
the elements annihilated by a power of $f$); and the iterated case is
the file \url{LocalizationOverlap}, which shows that the localisation at $g$ of a
localisation at $f$ is a localisation at $fg$. Its 44 declarations open
the universal property, the algebra maps and their own definitions, and
no definition of the construction (0.16~s). Univalence, available
here, does not remove this work: to identify two localisations one must
still construct the comparison isomorphism.

\paragraph{Human checking.} Ganesalingam and Gowers~\cite{GG} build a
prover whose output reads as a human proof; in it, expanding a definition
is a move of its own, of low priority, taken when nothing else applies.
Ganesalingam~\cite{Gan} studies the language of mathematical texts,
including definitions. Empirical studies of how mathematicians check
proofs~\cite{Weber,IA} find that experts check steps locally, each
against its justification, and spend their time on the steps whose
justification is implicit. We know of no study of when a human reader
opens a definition; the measurement proposed here gives a prediction
that such a study could test.

% ======================================================================
\section{Limitations and open questions}\label{sec:open}

The results are preliminary: the classification of the steps of a text
is done by hand, and the declarations of the model that open the most
have not yet been read against their texts.

We do not give a soundness or completeness proof of the conversion
algorithm. Reference~\cite{BCDE} supplies the universe discipline;
definitions follow $\Lambda\Delta$. The labelled inductive types with
primitive recursion, nominal records, fixed fields and module rules
described here are additional rules, not all part of \cite{BCDE}. Their
combined metatheory remains to be supplied. Other open points include
$\eta$ for records and the restrictions on local units (a local block
cannot introduce a data type or a module).

The checker, the examples and the developments are available at
\url{https://github.com/coquand/unit}. The implementation is about 5700
lines of Haskell:
\begin{center}
\begin{tabular}{lr}
\toprule
part & lines\\
\midrule
syntax, values, evaluation, conversion, checking of units and segments & 1850\\
omitted arguments: matching and writing back & 980\\
universe levels: the decision of constraints~\cite{BC} & 170\\
parser and resolver & 1740\\
printer, with the read-back of values to terms & 170\\
strict positivity & 130\\
driver, checking server and reports & 620\\
\bottomrule
\end{tabular}
\end{center}
Besides the
Haskell base library, the checker uses only the standard packages
\kw{containers} (finite maps and sets) and, for the checking server,
\kw{directory}; the parser is written by hand, without a parsing
library. Each declaration has a budget of ten seconds; a
checking server loads a development once and rechecks only the changed
files and those depending on them.

\section{Conclusion}

\sys\ is a type checker of about 5700 lines for a dependent type theory
with modules in the tradition of \textsc{Automath}, with one treatment of
definitions and nothing guessed. It has checked a model of univalent
type theory, synthetic algebraic geometry (schemes, the \'etale topology,
Severi--Brauer varieties),
synthetic Stone duality with the cohomology of compact Hausdorff
spaces, qcqs schemes with their functors of points, and Zariski's main
theorem with the local structure of \'etale algebras. These developments
are constructive, but the system is not: classical principles are
parameters like any other hypothesis, as in the simplicial set model.
A small, uniform system is therefore enough for substantial
mathematics, and its uniformity has a use of its own: the work of
checking can be observed and compared with the informal proof. The
extra notions of current proof assistants should be justified against
this baseline, not assumed, the more so as formal proofs come to be
written with the help of AI.

The system could be made smaller still, as de Bruijn argued in his plea
for weaker frameworks~\cite{Plea}: without data types and recursive
functions, the natural numbers and other inductive notions being
primitive notions and axioms declared in a context, as in
\textsc{Automath}, and with a simpler system of universes. This has a
cost: what a \kw{split} now computes, such as $\kw{plus}\ n\ \kw{zero} =
n$, would become a lemma, used by transport, as in objective type
theory~\cite{OTT}. Where to draw this line is the question of this
paper: how much computation a checker should perform silently.

So simplified, \sys\ would be close to Carneiro's MM0~\cite{MM0}, which
also aims at a small, efficient verifier for a weak framework: the
object logic is given by declared axioms, and definitions are unfolded
by a convertibility judgement of the framework, separate from the
assertions of the object logic. MM0 separates a readable specification
of axioms, definitions and statements from a proof certificate, produced
by the MM1 front end, whose conversion steps, congruence and unfolding
included, are explicit and can be shared. MM0 therefore already makes
unfolding explicit and inspectable. The difference is where the
unfolding steps are supplied: MM0 checks those written in a certificate;
\sys\ chooses them while checking a dependent proof term, by its uniform
strategy, and reports them, so that the work on the written proof can be
related to the reading of an accompanying text.

\section*{Acknowledgements}

The author supplied the design of \sys\ and directed its development
and the interpretation of the checking experiments. Claude (Anthropic)
produced the implementation code, and Claude and Codex (OpenAI)
produced the formalisation code through iterative work under that
direction. This attribution concerns code production; the mathematical
sources are cited in the text. The resulting text and software remain
the author's responsibility.


\begin{thebibliography}{99}
\bibitem{ACL} M.~E.~Alonso, T.~Coquand and H.~Lombardi. Revisiting
  Zariski main theorem from a constructive point of view. \emph{Journal
  of Algebra} 406 (2014), 46--68.
\bibitem{Zeu} M.~Zeuner. Univalent foundations of constructive
  algebraic geometry. Stockholm University, 2024.
\bibitem{ZH} M.~Zeuner and M.~Hutzler. The functor of points approach
  to schemes in Cubical Agda. 2024.
\bibitem{BC} M.~Bezem and T.~Coquand. Loop-checking and the uniform word
  problem for join-semilattices with an inflationary endomorphism.
  \emph{Theoretical Computer Science} 913, 2022.
\bibitem{BCDE} M.~Bezem, T.~Coquand, P.~Dybjer and M.~Escard\'o. Type
  theory with explicit universe polymorphism. arXiv:2212.03284.
\bibitem{OTT} B.~van~den~Berg and M.~den~Besten. Quadratic type checking
  for objective type theory. arXiv:2102.00905, 2021.
\bibitem{BCH} A.~Bastounis, F.~Circelli and A.~C.~Hansen. Navier--Stokes
  lost in translation: why Lean verification of AI autoformalisation does
  not guarantee correct natural language proofs. arXiv:2610.08144, 2026.
\bibitem{Burstall} R.~M.~Burstall. Inductively defined functions in functional
  programming languages. \emph{Journal of Computer and System Sciences}
  34(2/3):409--421, 1987.
\bibitem{SiL} K.~Buzzard, C.~Hughes, K.~Lau, A.~Livingston,
  R.~Fern\'andez~Mir and S.~Morrison. Schemes in Lean.
  \emph{Experimental Mathematics} 31(2):355--363, 2022.
  arXiv:2101.02602.
\bibitem{C96} T.~Coquand. An algorithm for type-checking dependent types.
  \emph{Science of Computer Programming} 26(1--3):167--177, 1996.
\bibitem{cubicaltt} C.~Cohen, T.~Coquand, S.~Huber and A.~M\"ortberg.
  Cubical type theory: a constructive interpretation of the univalence
  axiom. In \emph{TYPES 2015}, LIPIcs 69, 2018. Implementation
  \texttt{cubicaltt}: \url{https://github.com/mortberg/cubicaltt}.
\bibitem{CL} T.~Coquand and H.~Lombardi. \emph{R\'esolutions libres
  finies. M\'ethodes constructives}. Manuscript, 2024.
\bibitem{CQ} T.~Coquand and C.~Quitt\'e. Constructive finite free
  resolutions. \emph{Manuscripta Mathematica} 137 (2012), 331--345.
\bibitem{CT} T.~Coquand and C.~T\^ete. An elementary proof of Wiebe's
  theorem. \emph{Journal of Algebra} 499 (2018), 103--110.
\bibitem{Ced} J.~Cederquist. \emph{A Pointfree Approach to Constructive
  Analysis in Type Theory}. PhD thesis, Chalmers, 1997.
\bibitem{MM0} M.~Carneiro. Metamath Zero: designing a theorem prover
  prover. In \emph{Intelligent Computer Mathematics (CICM 2020)},
  LNCS 12236, Springer, 2020, 71--88.
  \href{https://doi.org/10.1007/978-3-030-53518-6_5}{doi:10.1007/978-3-030-53518-6\_5}.
\bibitem{Rem} T.~Coquand. Some remarks about dependent type theory. In
  B.~Meyer, editor, \emph{The French School of Programming}, Springer,
  2023, 175--202.
  \href{https://doi.org/10.1007/978-3-031-34518-0_8}{doi:10.1007/978-3-031-34518-0\_8}.
\bibitem{LD} N.~G.~de~Bruijn. Generalizing Automath by means of a
  lambda-typed lambda calculus. In \emph{Mathematical Logic and
  Theoretical Computer Science}, Lecture Notes in Pure and Applied
  Mathematics 106, Marcel Dekker, 1987, 71--92. Reprinted in
  \emph{Selected Papers on Automath}, North-Holland, 1994.
\bibitem{MV} N.~G.~de~Bruijn. The mathematical vernacular, a language
  for mathematics with typed sets. In \emph{Workshop on Programming
  Logic}, Marstrand, 1987. Reprinted in \emph{Selected Papers on
  Automath}, North-Holland, 1994.
\bibitem{Plea} N.~G.~de~Bruijn. A plea for weaker frameworks. In
  G.~Huet and G.~Plotkin (eds.), \emph{Logical Frameworks}, Cambridge
  University Press, 1991, 40--67.
\bibitem{Survey} N.~G.~de~Bruijn. A survey of the project Automath. In
  \emph{To H.~B.~Curry: Essays on Combinatory Logic, Lambda Calculus and
  Formalism}, Academic Press, 1980, 579--606.
\bibitem{TT} M.~H.~Escard\'o and contributors. \textsc{TypeTopology}: univalent mathematics in Agda.
  \url{https://github.com/martinescardo/TypeTopology}.
\bibitem{Lean4} L.~de~Moura and S.~Ullrich. The Lean 4 theorem prover and
  programming language. In \emph{CADE-28}, LNCS 12699, Springer, 2021.
\bibitem{Frege} G.~Frege. \emph{Begriffsschrift, eine der arithmetischen
  nachgebildete Formelsprache des reinen Denkens}. Nebert, Halle, 1879.
  English translation in J.~van Heijenoort, editor, \emph{From Frege to
  G\"odel}, Harvard University Press, 1967.
\bibitem{GG} M.~Ganesalingam and W.~T.~Gowers. A fully automatic theorem
  prover with human-style output. \emph{Journal of Automated Reasoning}
  58, 2017, 253--291.
\bibitem{Gan} M.~Ganesalingam. \emph{The Language of Mathematics}. LNCS
  7805, Springer, 2013.
\bibitem{CU} D.~Gratzer, J.~Sterling, C.~Angiuli, T.~Coquand and
  L.~Birkedal. Controlling unfolding in type theory. arXiv:2210.05420,
  2022.
\bibitem{IA} M.~Inglis and L.~Alcock. Expert and novice approaches to
  reading mathematical proofs. \emph{Journal for Research in Mathematics
  Education} 43, 2012, 358--390.
\bibitem{Lamport} L.~Lamport. How to write a proof. \emph{American
  Mathematical Monthly} 102, 1995, 600--608.
\bibitem{SSD} F.~Cherubini, T.~Coquand, F.~Geerligs and H.~Moeneclaey.
  A foundation for synthetic Stone duality. arXiv:2412.03203, 2024.
\bibitem{CCA} F.~Cherubini, T.~Coquand, F.~Geerligs and H.~Moeneclaey.
  Cohomology in synthetic Stone duality. Manuscript, 2026.
\bibitem{SAG} F.~Cherubini, T.~Coquand and M.~Hutzler. A foundation for
  synthetic algebraic geometry. arXiv:2307.00073.
\bibitem{Wang} Q.~Wang. \emph{A Haskell Implementation for a Dependent
  Type Theory with Definitions}. Master's thesis, Chalmers, 2021.
\bibitem{Weber} K.~Weber. How mathematicians determine if an argument is
  a valid proof. \emph{Journal for Research in Mathematics Education} 39,
  2008, 431--459.
\bibitem{Isar} M.~Wenzel. Isar: a generic interpretative approach to
  readable formal proof documents. In \emph{TPHOLs'99}, LNCS 1690,
  Springer, 1999.
\end{thebibliography}
\end{document}